\documentclass[aps,10pt,prl,twocolumn,superscriptaddress,noeprint,longbibliography,floatfix]{revtex4-2}
\PassOptionsToPackage{%
  pdfstartview=FitH,
  breaklinks=true,
  bookmarks=true,
  colorlinks=true,
  anchorcolor=black,
  citecolor=blue,
  filecolor=black,
  menucolor=black,
  urlcolor=blue,
  linkcolor=blue
}{hyperref}
\usepackage{CJK}
\usepackage{physics}
\usepackage{soul}
\usepackage[caption=false,subrefformat=parens,labelformat=parens]{subfig}
\usepackage{color} 
\usepackage{placeins}
\usepackage{orcidlink} 
\usepackage[table,dvipsnames]{xcolor}
\usepackage{wasysym} %
\usepackage{graphicx}
\usepackage{amssymb}
\usepackage{multirow} 
\usepackage{booktabs} %

\usepackage{comment}

\usepackage[margin=0.8in]{geometry}
\usepackage{tikz}
\usetikzlibrary[arrows.meta,positioning]
\usepackage{xcolor}
\usepackage{lipsum}

\definecolor{arrowred}{RGB}{190,45,45}
\definecolor{arrowblue}{RGB}{35,85,175}

\newcommand{\triangleschematic}[3]{%
  \begin{tikzpicture}[
      x=0.78cm,y=0.78cm,
      site/.style={circle,draw=black,fill=white,line width=1.1pt,inner sep=1.7pt},
      bond/.style={draw=black,line width=1.0pt},
      charge/.style={circle,draw=black,fill=white,line width=0.7pt,
                     minimum size=4mm,inner sep=0pt,font=\bfseries\small}
    ]
    \begin{scope}[scale=0.9]
    \coordinate (Lbase) at (0,1.3);
    \coordinate (Lshared) at (1.5,1.3);
    \coordinate (Ltop) at (0.75,2.599);
    \coordinate (Cleft) at (1.5,1.3);
    \coordinate (Cright) at (3.0,1.3);
    \coordinate (Cbottom) at (2.25,0);
    \coordinate (Rshared) at (3.0,1.3);
    \coordinate (Rbase) at (4.5,1.3);
    \coordinate (Rtop) at (3.75,2.599);
    \coordinate (center) at (2.25,0.866);

    \draw[bond] (Lbase)--(Lshared)--(Ltop)--cycle;
    \draw[bond] (Cleft)--(Cright)--(Cbottom)--cycle;
    \draw[bond] (Rshared)--(Rbase)--(Rtop)--cycle;

    \draw[bond,dashed] (center)--(Cleft) (center)--(Cright) (center)--(Cbottom);
    \foreach \p in {Lbase,Lshared,Ltop,Cright,Cbottom,Rbase,Rtop,center}{\node[site] at (\p) {};}
    \node[charge] at (0.75,1.72) {$#1$};
    \node[charge] at (3.75,1.72) {$#2$};

    \pgfmathsetmacro{\arrowscale}{1.0}
    
   \draw[-{latex[length=2.1mm]},arrowred,line width=1.7pt]
     (Lshared)++(-30:{-0.5*\arrowscale*#3}) -- ++(-30:{\arrowscale * 1.1 * #3});
   \draw[-{latex[length=2.1mm]},arrowblue,line width=1.7pt]
     (Rshared)++(30:{-0.5*\arrowscale*#3}) -- ++(30:{\arrowscale *1.1* #3});
     
    \end{scope}
  \end{tikzpicture}%
}
\usepackage{siunitx}
\usepackage{mhchem}

\newcommand{\nwalker}{n_{\rm walker}}

\newcommand{\beq}{\begin{equation}}
\newcommand{\eeq}{\end{equation}}

\renewcommand\[{\begin{equation}}
\renewcommand\]{\end{equation}}

\newcommand{\figref}[2]{\hyperref[#1]{\autoref*{#1}(#2)}}

\graphicspath{{FIG_arxiv_v1/}}

\AtBeginDocument{
	\heavyrulewidth=.08em
	\lightrulewidth=.05em
	\cmidrulewidth=.03em
	\belowrulesep=.65ex
	\belowbottomsep=0pt
	\aboverulesep=.4ex
	\abovetopsep=0pt
	\cmidrulesep=\doublerulesep
	\cmidrulekern=.5em
	\defaultaddspace=.5em
}%

\makeindex

\newcommand{\thistitle}{Defect Poisoning of Quantum Spin Ice}

\begin{document} 

\title{\thistitle}

\author{Alaric L. Sanders
\orcidlink{0000-0003-4283-0566}}
\affiliation{Helmholtz-Zentrum Berlin f\"ur Materialien und Energie, 
14109 Berlin, Germany}
\affiliation{Dahlem Center for Complex Quantum Systems and Fachbereich Physik, Freie Universit\"at Berlin, 
14195 Berlin, Germany}
\affiliation{T.C.M. Group, University of Cambridge, Cambridge, CB3 0US, UK}
\author{Gautam K. Naik \orcidlink{0009-0004-2379-5026}}
\affiliation{Department of Physics, Boston University, Boston, Massachusetts, 02215, USA}
\author{Jonathan N. Hall\'en \orcidlink{0000-0003-4883-4832}}
\affiliation{Department of Physics, Boston University, Boston, Massachusetts, 02215, USA}
\affiliation{Department of Physics, Harvard University, Cambridge, Massachusetts 02138, USA}
\author{Robin Sch\"afer \orcidlink{0000-0001-9728-2371}
\thanks{\href{mailto:robin_schaefer@fas.harvard.edu}{robin\_schaefer@fas.harvard.edu}}}
\affiliation{Department of Physics, Harvard University, Cambridge, Massachusetts 02138, USA}

\begin{abstract}
The hunt for a material realization of quantum spin ice has motivated more than two decades of experimental effort.
The candidate materials inevitably contain crystal imperfections, such as magnetic vacancies, whose effects are often disregarded.
Here, we show that experimentally relevant levels of dilution can qualitatively reshape the low-energy behavior, as nearby vacancies generate quantum fluctuations that are absent in the clean system.
Already at dilution levels as low as two percent, well below those reported in cerium-based pyrochlores, these vacancy-induced processes connect percolating clusters of spins and dominate over the conventional quantum-spin-ice dynamics.
We therefore argue that magnetic vacancies in current experiments can strongly contaminate, and potentially completely obscure, the sought-after signatures of quantum spin ice.
We support these conclusions using large-scale, unbiased quantum Monte Carlo simulations and exact diagonalization.

\end{abstract}
\maketitle

As Wolfgang Pauli supposedly once remarked, ``Festk\"orperphysik ist eine Schmutzphysik'' (``solid-state physics is the physics of dirt''), a phrase that aptly captures the ubiquity of disorder and imperfections in real solid-state systems~\cite{natelson_2018}.
Although remarkable strides have been made toward producing ultrapure crystals---defect densities in silicon can be measured in parts per trillion~\cite{silicon_2014}---chemical imperfections remain an unavoidable feature of most solid-state experiments.
Many phases of matter are robust against such imperfections. 
A prominent example is topological order, where quantum mechanics produces phases remarkably robust against disorder and local perturbations, most famously in the quantum Hall effect~\cite{tsui_two_1982,wen_ground_1990,bravyi_topological_2010}.
Other phases, however, do not benefit from such protection, and imperfections can destabilize the phase or, in some cases, drive transitions to qualitatively different states of matter~\cite{binder_spin_1986,henly_order_1989,evers_anderson_2008,freysoldt_first_2014}.

\begin{figure*}
    \centering
    \includegraphics{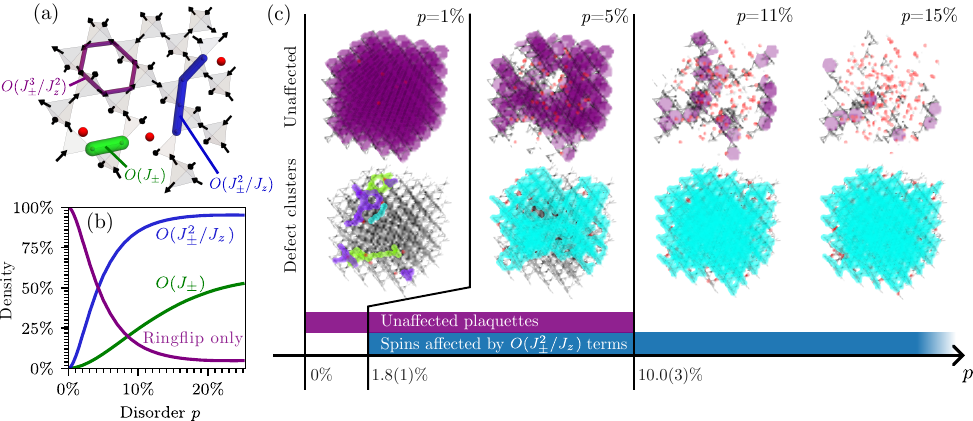}
    \caption{(a) Pyrochlore lattice with spins in a ``two-in-two-out'' ice configuration. 
    It illustrates defect-induced perturbative processes of order $O(J_\pm)$ (green) and $O\left(J_\pm^2/J_z\right)$ (blue) between the missing spins (red spheres). 
    The conventional ring-exchange process of order $O\left(J_\pm^3/J_z^2\right)$ is shown as a purple hexagon.
    (b) Fraction of spins affected by first-order processes (green) and by either first- or second-order processes (blue).
    The purple curve shows the fraction of spins whose dynamics are governed exclusively by third-order ring exchange.
    (c) Illustration of the connected networks formed by clean, unaffected hexagons (purple, top row), and by spins affected by lower-order processes (bottom row) for several defect densities $p$.
    Different colors denote disconnected clusters of $O(J_\pm^2/J_z)$ spins.
    The colored bars below indicate the percolation regimes of connected, unaffected plaquettes (top) and spins affected by lower-order processes (bottom).
    }\label{fig:1}
\end{figure*}

Disorder is one possible reason why a definitive realization of a $U(1)$ quantum spin liquid has eluded decades of experimental effort.
The fragility of quantum spin liquids can be understood from the mechanism by which many of them arise.
When an extensively degenerate classical ground state manifold is subjected to quantum fluctuations, the degeneracy can be lifted in favor of a strongly correlated, entangled ground state~\cite{lhuillier_introduction_2011,chalker_geometrically_2011,chalker_spin_2014}.
Disorder competes directly with this mechanism: it can partially or completely lift the underlying classical degeneracy or introduce additional local quantum fluctuations that are absent in the clean system. 
As a result, even weak disorder can qualitatively alter the low-energy physics and potentially destabilize the quantum spin-liquid state.

In this Letter, we consider the $U(1)$ quantum spin liquid model on the pyrochlore lattice known as quantum spin ice (QSI), which realizes emergent quantum electrodynamics~\cite{hermele_pyrochlore_2004,savary_coulombic_2012,shannon_quantum_2012,benton_seeing_2012, udagawa_spin_2021, gingras_quantum_2014}. 
Recent efforts to realize QSI have focused on Ce-based pyrochlores, in which magnetic Ce ions form a pyrochlore lattice of Kramers doublets~\cite{sibille_candidate_2015,gaudet_quantum_2019,rau_frustrated_2019,gao_experimental_2019,sibille_quantum_2020,bhardwaj_sleuthing_2022,smith_case_2022,yahne_dipolar_2024,smith_single_2025,smith_two-peak_2025,gao_neutron_2025}.
One of the most prominent sources of disorder in these materials is the oxidation of Ce$^{3+}$ ions to non-magnetic Ce$^{4+}$, which effectively dilutes the magnetic lattice~\footnote{Stuffing provides another important source of disorder, in which Ce ions exchange positions with transition-metal ions on the interpenetrating sublattice. This likewise leaves a non-magnetic site on the primary magnetic lattice, while potentially introducing an additional magnetic degree of freedom on the interpenetrating pyrochlore lattice.}.
For current Ce$_2$\emph{B}$_2$O$_7$ (\emph{B} = Zr, Hf) crystals, the concentration of non-magnetic Ce$^{4+}$ ions is estimated to be below approximately $5\%$ for specific-heat measurements and below $10\%$ for single-crystal neutron-scattering experiments~\cite{gao_experimental_2019,gaudet_quantum_2019,yahne_dipolar_2024}.
We model this disorder by randomly removing spins from the lattice and show that dilution at the few-percent level introduces previously unaccounted-for processes that qualitatively alter the low-energy physics.

In the canonical, clean QSI model, virtual spin-flip processes generate an effective ring-exchange interaction around hexagonal loops of the pyrochlore lattice at third order in perturbation theory \cite{hermele_pyrochlore_2004}.
Introducing non-magnetic vacancies opens additional pathways for quantum fluctuations that arise already at lower orders in perturbation theory.
When two vacancies are separated by two (four) lattice sites, a first- (second-) order process emerges (see \figref{fig:1}{a}).
These defect-induced processes rapidly involve a large fraction of the lattice as the dilution is increased.
In \figref{fig:1}{b}, we show the fraction of spins participating in first-order processes in green and in either first- or second-order processes in blue.
Remarkably, at dilution levels of only a few percent, more than half the spins participate in these lower-order processes.
At dilution levels as low as $2\%$, the spins participating in first- or second-order quantum processes form extensive, percolating clusters.
These percolating networks of lower-order quantum dynamics will leave extensive signatures in experimentally accessible observables, including the specific heat and equal-time structure factor.
Our results, therefore, challenge recent experimental interpretations and call into question the stability of QSI in currently available crystals.

\paragraph{Microscopic model and perturbation theory.}
We begin with the well-understood case of disorder-free QSI~\cite{hermele_pyrochlore_2004,savary_coulombic_2012,shannon_quantum_2012,benton_seeing_2012, udagawa_spin_2021, gingras_quantum_2014}.
The generic nearest-neighbor model is
\begin{align}
    H = \sum_{\langle ij\rangle} J_x\,S^x_i S^x_j + J_y\,S^y_i S^y_j + J_z\,S^z_i S^z_j\,,\label{eq:H}
\end{align}
where $S^\alpha_i$ are spin-1/2 operators placed at the vertices of a pyrochlore lattice, with local quantization axes $z$ \cite{bramwell_spin_2001}. 
The sum runs over all nearest-neighbor pairs.
The low-energy theory giving rise to the $U(1)$ liquid emerges in the limit where one coupling is antiferromagnetic and dominant, e.g.\ $J_z \gg |J_x|,\,|J_y|$.
In the Ising limit ($J_x=J_y=0$), any configuration with two spins pointing in and two pointing out of every tetrahedron minimizes the energy. 
This local constraint does not select a unique global ground state; it is satisfied by an extensive number of degenerate ``ice states''~\cite{harris_geometrical_1997,ramirez_zero-point_1999,pauling_structure_1935,anderson_ordering_1956}.
Violations of this two-in-two-out constraint are referred to as magnetic monopoles, and carry an energy penalty of order $O(J_z)$.
In the clean, non-diluted model, small off-diagonal terms mediate transitions within perturbation theory between these classical ground states by flipping closed loops of head-to-tail spins.
The smallest such loops are hexagons, and away from the Ising limit they define the effective ring-exchange Hamiltonian that lifts the classical degeneracy,
\begin{align}
    H_{\mathrm{ring}} = \frac{12\,J_\pm^3}{J_z^2}\sum_{\hexagon}
    \left( S^+_1 S^-_2 S^+_3 S^-_4 S^+_5 S^-_6 + \mathrm{h.c.}\right),
    \label{eq:RingExchange}
\end{align}
where the sum runs over all hexagonal plaquettes~\cite{hermele_pyrochlore_2004}.
For simplicity, we set $J_x=J_y$ and define $J_\pm=\frac{1}{2}J_{x/y}$.
These hexagon contributions arise at third order in perturbation theory, leaving the system with two well-separated energy scales, $O(J_z)$ and $O(J_\pm^3/J_z^2)$.

Classical spin ice ($J_\pm=0$) is robust to high levels of random dilution~\cite{lin_nonmonotonic_2014,sen2013coulomb, sen2015topological}.
Removing a single spin leaves two incomplete tetrahedra that can no longer satisfy the two-in-two-out rule.
Instead, they adopt one-in-two-out or two-in-one-out configurations, which can be viewed as bound monopoles.
These can interact with nearby free monopoles~\cite{petrova_hydrogenic_2015}, but cannot be frozen out at any temperature. 
Nevertheless, in the ultra-low dilution limit, where all vacancies are far apart, the physics at energy scales well below $J_z$ remains governed by $H_{\rm ring}$, now operating within a modified set of ice states with isolated bound monopoles.

The picture changes drastically with increasing dilution, as vacancies occur within a few lattice sites of one another and lower-order terms in perturbation theory enter.
These unlock new energy scales of order $J_\pm$ and $J_\pm^2/J_z$.  
To see this, consider the case of two bound monopoles separated by two spins aligned head-to-tail.
Flipping these spins inverts the bound monopole charges and thereby connects two classical ice states as illustrated here:\\

\noindent
\makebox[\columnwidth][s]{%
    ~~\triangleschematic{+}{-}{1}\hfill
    \raisebox{0.85cm}{\Huge$\boldsymbol{\leftrightarrow}$}\hfill
    \triangleschematic{-}{+}{-1}~~
}

\noindent The left and right tetrahedra are incomplete, leaving triangles with bound monopoles.
The intermediate tetrahedron remains intact.
Formally, terms of the form $J_\pm (S^+_i S^-_j + S^-_iS^+_j)$ survive a perturbative Schrieffer-Wolff projection onto the ice manifold~\footnote{Note that if a tetrahedron has two missing spins it does not host bound monopoles, and two- or four-spin terms ending at such tetrahedra are therefore not allowed.}.
Similarly, bound monopoles separated by a string of four spins allow terms at second order in perturbation theory, giving rise to an energy scale of order $O(J_\pm^2/J_z)$.

The new energy scales give rise to a hierarchy in the limit $J_z \gg |J_\pm|$: at high temperatures $T \gg |J_\pm|$, the system behaves classically and can be modeled by diffusive and bound monopoles.
Upon cooling to $T \sim |J_\pm|$, the physics is confined to the ice manifold, and disorder-induced first-order processes become relevant.
Bound monopole states separated by two sites want to hybridize, lowering the overall energy by $O\left(J_\pm\right)$ as they drop into their effective ground state $(\ket{\uparrow \downarrow} \pm \ket{\downarrow\uparrow}) / \sqrt{2}$.

At $T \sim  J_\pm^2/J_z $, first-order processes are frozen out \footnote{Note that first-order networks may be frustrated, leading to finite residual entropy.}, and second-order processes enter, similarly hybridizing four-spin chains.
Only at $T \sim |J_\pm|^3/J_z^2$ can one expect to find signatures of the celebrated emergent quantum electrodynamics.
Even then, ring exchange competes with new, non-hexagonal third-order processes connecting bound monopoles across six sites.

\begin{figure}
    \centering
    \includegraphics{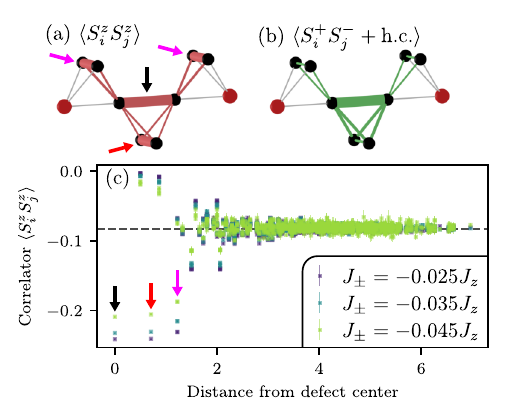}
    \caption{Green's function Monte Carlo simulation of a cluster with $N=432-2$ sites containing two magnetic vacancies (red spheres) separated by two lattice sites, which support a first-order process.
    Panels (a,b) show, respectively, the $\langle S^z_i S^z_j\rangle$ and $\langle S^+_i S^-_j+\text{h.c.}\rangle$ correlations at $J_\pm/J_z=-0.045$.
    The bond that supports the first-order process is indicated by a black arrow, and the apical bonds are indicated by purple and red arrows. 
    Thickness relative to the central bond is proportional to the absolute value of the corresponding correlator, while red (green) bonds indicate negative (positive) correlations. 
    Panel (c) shows the distribution of nearest-neighbor correlators $\langle S^z_i S^z_j\rangle$ as a function of distance from the central dimer (black arrow) for different values of $|J_\pm|/J_z$, in units of the spin-spin bond length.
    }
    \label{fig:2}
\end{figure}

The hierarchy laid out above is well-defined only in the limit of infinitesimal $\vert J_\pm \vert /J_z$; for finite $J_\pm$, the scales cannot be cleanly separated.
To demonstrate that dimer formation persists away from the ideal limit, we perform unbiased numerical simulations at finite $|J_\pm|/J_z$.
Specifically, we compute the ground state of \autoref{eq:H} for a system with $N=432-2$ sites using Green's function Monte Carlo (GFMC)~\cite{ceperley_ground_state_1980,trivedi_ground-state_1990,sorella_green_1998,calandra_numerical_1998,becca_quantum_2017,PyGFMC,supplemental_material}, where two vacancies are separated by two lattice sites.
Although this technique can only be applied in the sign-problem free regime $J_\pm\le0$, we expect qualitatively similar behavior when $J_\pm>0$.
\figref{fig:2}{a,b} show the nearest-neighbor expectation values $\langle S^z_i S^z_j\rangle$ and $\langle S^+_i S^-_j+\mathrm{h.c.}\rangle$ at $J_\pm/Jz=-0.045$ in the vicinity of the defects.
The corresponding clean system lies in the $U(1)_0$ spin-liquid phase, known to transition into an ordered state around $J_\pm/Jz\approx-0.052$~\cite{banerjee_unusual_2008}.
The connecting bond which supports the first-order process is strongly correlated, with $\langle S^z_i S^z_j\rangle\approx-0.21$ and $\langle S^+_i S^-_j+\mathrm{h.c.}\rangle\approx0.83$, indicating that they form a dimer state as expected.
Forming this dimer without violating ice rules on the intermediate tetrahedra and triangles places constraints on outer spin pairs.
These constraints manifest as strong $\langle S^z_i S^z_j\rangle$ correlations on the bonds apical to the central dimer.
As a result, the bonds connecting the apical spins to the central dimer exhibit substantially weaker $\langle S^z_i S^z_j\rangle$ correlations.
The transverse nearest-neighbor correlators, $\langle S^+_i S^-_j+\mathrm{h.c.}\rangle$, display a qualitatively different spatial pattern.
In particular, the apical bonds do not show any substantial increase in the transverse correlator, as applying a $S^+_i S^-_j$ term there creates monopoles on surrounding tetrahedra (not shown).

In \figref{fig:2}{c}, we show the distribution of the nearest-neighbor correlators $\langle S^z_i S^z_j\rangle$ as a function of distance from the central dimer for different values of $|J_\pm|/J_z$.
Defect-induced correlations remain confined to the immediate vicinity of the missing spins, while pairs farther from the structure shown in \figref{fig:2}{a,b} appear to remain disordered, retaining the clean-system value $\langle S^z_i S^z_j\rangle\approx-0.083$.
Correlations become increasingly pronounced as $|J_\pm|/J_z\rightarrow0$, in accordance with perturbation theory.
In the limit $|J_\pm|/J_z\rightarrow0$, we expect the two central spins to decouple from the remainder of the system, forming the product state $\ket{\psi} \otimes (\ket{\uparrow\downarrow}+\ket{\downarrow\uparrow})/\sqrt{2}$, where $\ket{\psi}$ is the many-body wavefunction of the remaining 428 spins (see SM~\cite{supplemental_material} for further details).

\paragraph*{Disordered networks.}
The bound monopoles do not only form pairs, but can also be part of large networks.
Because of the high connectivity of the pyrochlore lattice---each site has 48 fourth-nearest neighbors (relevant for first-order processes) and 126 sixth-nearest neighbors (relevant for second-order processes)~\footnote{The pyrochlore has 240 eighth-nearest neighbors which are relevant for third-order processes.}---large networks of connected bound monopoles form at only a few percent dilution.
As the dilution level is increased, these networks undergo a percolation transition, at which point they span the system and involve a non-vanishing fraction of the spins.
For first-order processes, the bound-monopole network percolates at around $10\%$ dilution and operates at an energy scale $O(J_\pm)$.
Including both first- and second-order processes lowers the percolation threshold to approximately $2\%$, with about half of all spins belonging to the network already at $5\%$ dilution.
At third order, where defect-induced processes occur on the same energy scale as the conventional ring exchange, the bound-monopole network percolates at only about $0.4\%$ dilution, with roughly half of the spins participating by $1\%$ dilution.

The defining feature of QSI is the presence of long-wavelength photons that propagate through the resonating ring-exchange terms of the effective Hamiltonian in \autoref{eq:RingExchange}.
This raises a natural question: up to what dilution level do hexagons unaffected by lower-order processes continue to percolate?
We define two clean hexagons as connected when they share at least one spin and observe a percolation transition at approximately $10\%$ dilution. 
Above 10\%, clean hexagons only form finite-size clusters that cannot support long-wavelength photons.

We visualize the different regimes that appear as a function of dilution in \figref{fig:1}{c}.
The upper row illustrates networks formed by the unaffected, clean plaquettes; the lower row highlights the network of spins affected by first- or second-order processes.
At 1\% dilution clean plaquettes cover most of the system and are therefore likely capable of supporting (dressed) long-wavelength photons.
In contrast, the affected spins form isolated islands in which the local physics is dominated by nearby defects.
At a defect density of 5\%, comparable to current Ce$_2$Zr$_2$O$_7$ crystal quality, the network of unaffected plaquettes is strongly perforated but remains percolating, potentially allowing long-wavelength photon excitations to persist.
At the same time, roughly half of the spins are affected by lower-order processes and form a percolating network of connected bound monopoles.
Since such processes affect approximately half the system, we expect them to produce pronounced signatures in any experimental measurements.
By 11\% dilution, approximately the defect density for neutron-scattering experiments, the clean plaquettes cease to percolate and instead form isolated regions.
These finite regions alone are not capable of supporting the characteristic long-wavelength photons of the clean system.
Most spins instead belong to the connected network of bound monopoles generated by lower-order processes.

\begin{figure}
    \centering
    \includegraphics[width=\columnwidth]{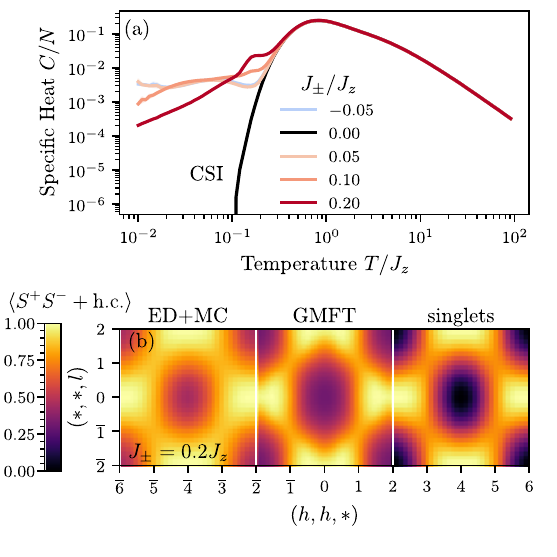}
    \caption{Observables of the first-order-only model obtained from 
    quantum-classical hybrid Monte Carlo (ED+MC)
    simulations. We use a $N\approx 8192(1-p)$ cluster at $p=5\rm\%$ disorder, averaged over 64 disorder realizations. 
    Panel (a) shows the heat capacity for several values of $J_\pm/J_z$, relative to the classical spin ice (CSI) result.
    Panel (b) shows the transverse $\langle S^+_iS^-_j+\mathrm{h.c.}\rangle$ static structure factor for $J_\pm=0.2J_z$, computed within different theoretical models.
    The leftmost third shows results from ED+MC at $T=0.01J_z$.
    The center third shows the same correlations within GMFT~\cite{desrochers_competing_2022,savary_coulombic_2012}, at $T=0$.
    The right third of panel (b) is obtained by placing two spins along all six bond orientations in their singlet state. All structure-factor plots are normalized to their respective maximum values.
    }
    \label{fig:3}
\end{figure}

\paragraph*{Experimental consequences.}
The quality of current samples is within the intermediate regime shown in the central columns in \figref{fig:1}{c}, where vast networks of bound monopoles are expected to leave experimentally measurable signatures across a broad range of energy scales.
To assess these effects, we employ a quantum-classical hybrid Monte Carlo method to approximate the first-order effective Hamiltonian.
The local networks supporting first-order quantum fluctuations are coupled to one another via a classical spin ice bulk, which is treated with classical Monte Carlo.
At sufficiently small $p$, each cluster can be exactly diagonalized~\cite{supplemental_material}.

The additional energy scale $J_\pm$ manifests in the specific heat.
Clean QSI exhibits two peaks in the specific heat: the Schottky anomaly at $T\sim J_z$ associated with the freezing-out of monopoles, and a second peak associated with the release of the residual ice entropy at $T\sim \lvert J_\pm\rvert^3/J_z^2$~\cite{kato_numerical_2015,huang_dynamics_2018}.
We show the specific heat as a function of temperature in \figref{fig:3}{a} at a dilution level of $p=5\%$, for which approximately $10\%$ of the spins participate in first-order processes.
While the signal of classical spin ice dies out at $T\sim 0.1J_z$, the first-order processes carry spectral weight to much lower temperature.
The magnitude of this feature is approximately one order of magnitude smaller than the Schottky anomaly, consistent with only about $10\%$ of the spins belonging to these first-order networks.
Notably, these calculations do not consider any effects from the second-order processes.
These involve half the spins in the system, and we expect they would contribute proportionately to the specific heat at $T\sim J_\pm^2/J_z$.
Several independent experiments on different Ce compounds~\cite{gao_experimental_2019,smith_case_2022,smith_two-peak_2025} have observed a broad hump in the specific heat, which cannot be captured by a clean nearest-neighbor model. 
This hints at the presence of a broad range of energy scales, which we propose could originate from the formation of defect networks of different sizes.

Neutron scattering is another important probe in the search for QSI, and it is therefore natural to ask whether defect-induced processes can obscure its characteristic signatures.
While we do not observe any visible changes in the $\langle S^z_iS^z_j\rangle$ correlations~\cite{supplemental_material}, the first-order processes manifest in the transverse structure factor, \figref{fig:3}{b}.
The leftmost panel shows data at $T=J_z/100$ and $J_\pm=J_z/5$, obtained using the hybrid Monte Carlo Ansatz.
Surprisingly, these results closely resemble the spinon correlations predicted for \textit{clean} QSI by gauge mean-field theory (GMFT)~\cite{desrochers_symmetry_2023,desrochers_competing_2022,savary_coulombic_2012} (middle panel).
Thus, the defect-induced transverse correlations mimic the signal expected for clean QSI, raising serious questions about the interpretation of neutron-scattering data without careful consideration of defects.
The observed pattern can be qualitatively reproduced by randomly placing \emph{isolated} dimers in their singlet ground state along the six possible bond orientations (right panel).

While the quality of current state-of-the-art experiments on Ce-based QSI candidates lies within the dilution regime we consider, their exchange parameters are generally not in the perturbative limit $\lvert J_\pm\rvert\ll J_z$.
For example, estimates for Ce$_2$Zr$_2$O$_7$ place it in the regime $J_x\approx J_y\gg J_z$~\cite{bhardwaj_sleuthing_2022,smith_case_2022}.
Nevertheless, we argue that clusters of nearby defects qualitatively alter the physics compared to the clean case and that strong local correlations persist beyond the perturbatively controlled regime (see SM~\cite{supplemental_material} for further details).
The spatially inhomogeneous correlation patterns induced by the defects may act as local pinning fields for valence-bond-crystal (VBC) phases that have been proposed as competing ground states away from the perturbative regime~\cite{hagymasi_possible_2021,astrakhantsev_broken-symmetry_2021,schafer_abundance_2022,hagymasi_enhanced_2022,pohle_ground_2023,cheng_closely_2026}.
However, randomly positioned defects may favor different VBC patterns and could therefore produce distinct domains rather than uniform long-range order.
We note that related spatially inhomogeneous patterns of enhanced bond correlations have also been found around non-magnetic vacancies in two-dimensional kagome antiferromagnets~\cite{dommange_static_2003,lauchli_static_2007}.

\paragraph*{Conclusion.}
Our results strongly suggest that dilution plays an important role in currently available samples. 
Firstly, a significant fraction of the system is dominated by lower-order processes whenever there is dilution at the parts-per-hundred scale. 
Sought-after QSI signatures must therefore be disentangled from this ``Schmutzphysik''.
Secondly, extensive, percolating networks of bound monopoles appear above 2\% dilution and rapidly involve the majority of the spins. 
A complete understanding of these random networks is beyond the scope of this work, but represents an intriguing problem of clear experimental relevance. 
If current samples do indeed have 5--10\% dilution as suggested~\cite{gao_experimental_2019,poree_crystal_2022,gaudet_quantum_2019}, their behavior is likely better understood in terms of these bound monopole networks than as a manifestation of clean QSI.

\vspace{0.5cm}
\begin{acknowledgments}
\paragraph*{Acknowledgments.}The authors thank Claudio Castelnovo, Bruce Gaulin, Chris Laumann, Roderich Moessner, Jeff Rau, Johannes Reuther, Nic Shannon, and Evan Smith for helpful discussions.
R.S. acknowledges support from the DFG under Project No. 575641691 and the Helmholtz-Zentrum Berlin. A.L.S. acknowledges support from the Helmholtz-Zentrum Berlin. Simulations used DanceQ~\cite{schaefer_danceq_high_2025, schaefer_codebase_release_2025}, NetKet~\cite{netket2:2019,netket3:2022}, and PyGFMC~\cite{PyGFMC}. 
J.N.H.\ acknowledges support from The Sweden-America Foundation.
Simulations were performed on the CURTA~\cite{curta2020} cluster at FU Berlin, the computing infrastructure of the Paderborn Center for Parallel Computing (PC2), and the Cannon cluster at Harvard University.
\end{acknowledgments}

\bibliography{references}

\appendix
\section{Supplementary Numerical Results}
\FloatBarrier
We provide further numerical evidence based on exact diagonalization, showing that the defect cluster disentangles from the system as $J_\pm/J_z \rightarrow 0$ and that the defect-dominated physics persists beyond the perturbatively controlled regime.
We consider the two-dimensional checkerboard lattice, also known as square ice, and a pyrochlore lattice.
\paragraph{Checkerboard lattice.} The checkerboard lattice implements corner-sharing ``tetrahedra'' in two dimensions by adding additional diagonal bonds with equal strength to every second plaquette on a square lattice.
As on the pyrochlore lattice, the Ising limit hosts an extensively degenerate manifold of ice states satisfying the two-in-two-out constraint on every ``tetrahedron,'' corresponding here to the crossed plaquettes.
This degeneracy is lifted at second order in perturbation theory by ring-exchange processes acting on the uncrossed plaquettes.
These processes flip configurations in which the spins are arranged head-to-tail around an uncrossed plaquette and have an amplitude of order $O\left(J_\pm^2/J_z\right)$.
Because the model is two-dimensional, it does not support a stable deconfined $U(1)$ Coulomb phase~\cite{polyakov_quark_1977}.
Instead, the gauge theory confines into a symmetry-breaking valence-bond crystal~\cite{fouet_planar_2001,henryOrderbyDisorderQuantumCoulomb2014,capponi_numerical_2017}.

\begin{figure}[t]
    \centering
    \includegraphics[width=\linewidth]{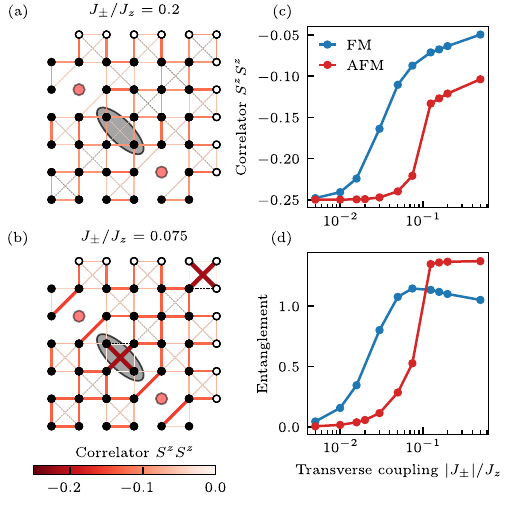}
    \caption{Exact diagonalization results for the checkerboard lattice with two defects (marked in red).
        Panels (a,b) show the nearest-neighbor correlations $|S^z_iS^z_j|$ for $J_\pm/J_z=0.2$ and $J_\pm/J_z=0.075$, respectively.
        The bond thickness and color encode the magnitude of the correlation.
        Markers highlighted in white are periodic images of the sites.
        Panel (c) shows the correlation on the central bond, highlighted in gray, for ferromagnetic (FM, $J_\pm<0$) and antiferromagnetic (AFM, $J_\pm>0$) transverse couplings.
        Panel (d) shows the von Neumann entanglement entropy between the central dimer and the remainder of the system.
        }
    \label{fig:4}
\end{figure}

We compute the ground state of a checkerboard lattice with two defects separated by two lattice sites (as in \autoref{fig:4}) using a system of $N=36-2$ spins, for several values of $J_\pm/J_z$.
Panels (a,b) of \autoref{fig:4} show the nearest-neighbor $S^z_iS^z_j$ correlations for different values of $J_\pm/J_z$.
The defects are marked in red, and the bond supporting the first-order process is highlighted in gray.
A second bond in the upper-right corner supports the same first-order process across the periodic boundary.
In panel (a), at $J_\pm/J_z=0.2$, no pronounced defect-induced correlations are visible.
Here, the valence-bond crystal dominates the ground state and produces an almost uniform correlation pattern.
In contrast, panel (b), at $J_\pm/J_z=0.075$, clearly exhibits a correlation pattern analogous to that observed in the GFMC calculation underlying \autoref{fig:2}.
The bond connecting the defects is strongly correlated, while the local ice constraints induce additional correlations on neighboring bonds.
This provides clear evidence that nearby defects locally destabilize the surrounding valence-bond state and generate the characteristic correlation pattern.
We expect the uniform bulk correlation pattern to be recovered sufficiently far from the defects.
Note that the same defect-induced pattern also appears in the upper-right corner, where the corresponding bond connects the two defects across the periodic boundary.

We further investigate the physics of the central bond in the limit $J_\pm/J_z\rightarrow 0$ in panels (c,d) of \autoref{fig:4} by examining its correlation and its entanglement with the remainder of the system.
Consistent with the patterns observed in panels (a,b), we identify a narrow crossover at which the plaquette phase is locally destabilized in favor of a modified correlation pattern between the defects.
The correlator decreases rapidly towards its maximally antiferromagnetic value around $\vert J_\pm\vert /J_z\approx 0.1$ as shown in \figref{fig:4}{c}.
Exact diagonalization also allows us to quantify the entanglement between the central dimer and the rest of the system.
In panel (d), we show the von Neumann entropy $S=-\operatorname{Tr}\left(\rho_{\mathrm{d}}\ln\rho_{\mathrm{d}}\right)$,
where $\rho_{\mathrm{d}}$ is the reduced density matrix of the two spins forming the central dimer.
The entropy closely tracks the behavior of the bond correlator and vanishes as $\lvert J_\pm\rvert/J_z\rightarrow 0$, indicating that the dimer becomes completely disentangled from the rest of the system.
This behavior is consistent with perturbation theory.

\paragraph{Pyrochlore lattice.} We also expect nearby defects to generate spatially inhomogeneous correlation patterns away from the perturbatively controlled regime in the pyrochlore lattice.
This can be understood by recasting the nearest-neighbor Hamiltonian in \autoref{eq:H} in terms of the square of the total (weighted) spin on each tetrahedron.
For arbitrary antiferromagnetic exchange couplings $J_x$, $J_y$, and $J_z$, the Hamiltonian can be written, up to an additive constant and an overall factor, as
\begin{equation}
    H = \sum_t \vec{J}_t^{\,2}\, , \qquad
    \vec{J}_t = \sum_{i \in t}
    \begin{pmatrix}
        \sqrt{J_x} S_i^x \\
        \sqrt{J_y} S_i^y \\
        \sqrt{J_z} S_i^z
    \end{pmatrix}
    \, .
\end{equation}
In contrast to the Ising case, it is not possible to minimize the energy of all tetrahedra simultaneously in the presence of transverse couplings.
However, this frustration is locally relieved in the presence of two nearby vacancies.
For example, in the geometry shown in \figref{fig:2}{a}, the energies of both triangular units and the intervening tetrahedra can be minimized together, leading to enhanced correlations on the bond connecting the defective tetrahedra.

\begin{figure}[t]
    \centering
    \includegraphics[width=\linewidth]{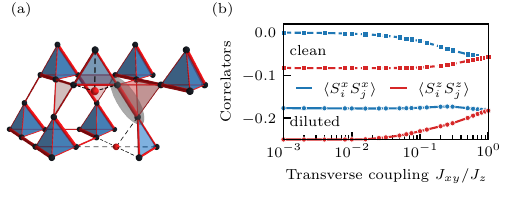}
    \caption{Exact diagonalization results for the pyrochlore antiferromagnet.
    Panel (a) shows the nearest-neighbor $S^zS^z$ correlations in the pyrochlore lattice constructed from a $2\times2\times2$ primitive unit cell, at the pure-Heisenberg $J_{xy}=J_z$ point. 
    Panel (b) shows $S^x_iS^x_j$ ($S^z_iS^z_j$) correlations on the bond highlighted in gray that connects the defective tetrahedra, as a function of the transverse coupling $J_{xy}=J_x=J_y$.
    The dashed lines refer to the clean $N=32$ system. 
        }
    \label{fig:5}
\end{figure}

We demonstrate this by performing exact diagonalization on the pyrochlore lattice constructed from $2\times2\times2$ primitive unit cells with two defects, as shown in \figref{fig:5}{a}.
The removed sites are marked in red, and we focus on the bond highlighted in gray, connecting the two triangles.

\figref{fig:5}{b} shows the $S_i^xS_j^x$ (blue) and $S_i^zS_j^z$ (red) correlators on the highlighted bond as functions of $J_{xy}=J_x=J_y$.
In the strict perturbative limit, at small $J_{xy}/J_z$, the $S^zS^z$ correlator approaches its maximally antiferromagnetic value because it is dominated by the lower-order processes in perturbation theory.
This is substantially enhanced compared to its clean-system value, around $-0.083$.
The transverse correlators most clearly distinguish the clean and diluted cases: the $S^xS^x$ correlations are strong only in the latter case, where the triangle charges may freely fluctuate.
The patterns remain robust as the transverse coupling is increased, with the central bond remaining much more strongly correlated than its environment throughout the parameter range considered.

The strength of the nearest-neighbor correlations at the Heisenberg point, $J_{xy}=J_z$, encoded by the bond thickness is shown in \figref{fig:5}{a}.
It clearly reveals an enhanced response in the intermediate tetrahedron between the defects, where both the connecting bond and the bond opposite to it are similarly enhanced.
This defect pattern induces a symmetry-breaking pattern across the entire system.
This is, however, not surprising for this finite lattice.
In addition to the resonating hexagonal loops that govern quantum spin ice, the small lattice constructed from $2\times2\times2$ primitive unit cells contains additional loops of length four that wind across the periodic boundaries.
The defects favor one of the states that exhibit strong correlations along these boundary loops (see Ref.~\cite{hagymasi_possible_2021}) and stabilize it.

\section{Green's Function Monte Carlo}
GFMC is a projection method onto the ground state of unfrustrated Hamiltonians~\cite{ceperley_ground_state_1980,trivedi_ground-state_1990,sorella_green_1998,calandra_numerical_1998,becca_quantum_2017}.
The algorithm computes diagonal and off-diagonal observables in the ground state by evolving a walker ensemble of fixed size $\nwalker$.
Within the continuous-time formalism, a branching step is applied after evolving for an imaginary-time interval $\beta$.
The observables are measured as a function of the total projection depth $\tau$, which is obtained by multiplying the number of branching steps by $\beta$.
Off-diagonal measurements are substantially more expensive than diagonal observables, as they require launching an auxiliary Markov chain.
Therefore, we used a smaller ensemble of $2^{15}$ walkers to compute the $\langle S^+_iS^-_j+\rm h.c.\rangle$ correlations, compared to $2^{17}$ walkers for $\langle S_i^zS_j^z\rangle$.
The data in \autoref{fig:2} of the main text are shown for $\tau=24$, obtained using step sizes of $\beta=1,\,0.6,\,0.6$ for $J_\pm/J_z=-0.025,\,-0.035,\,-0.045$, respectively.
The mean values and error bars are obtained from ten independent samples, each using $10\,000$ measurements for the off-diagonal observables and $200\,000$ measurements for the diagonal observables.
To reduce correlation between off-diagonal measurements, we launched an auxiliary chain only every 16th branching step.
The simulations were conducted using PyGFMC~\cite{PyGFMC}.

\begin{figure}[t]
    \centering
    \includegraphics[width=\linewidth]{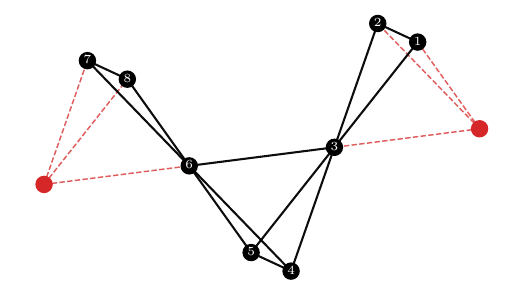}
    \caption{Illustration of the central eight-site structure.
    We use the following automorphism: two-site exchanges on the apex bonds, $(1,2)$, $(4,5)$, and $(7,8)$, as well as a reflection that exchanges the two triangles, $(1,8)(2,7)(3,6)$.
        }
    \label{fig:patch}
\end{figure}

We employ a guiding wavefunction that significantly improves the convergence of the simulation.
It provides the log-amplitude $\log\Psi(\ket{x})$ of a given spin configuration $\ket{x}=\ket{x_{\rm center}}\ket{x_{\rm bulk}}$, where $\ket{x_{\rm center}}$ contains the eight central spins shown in \autoref{fig:patch} (and \autoref{fig:2} of the main text), while $\ket{x_{\rm bulk}}$ contains the remaining bulk spins.
It is separated into two parts:
\begin{align}
\log \Psi(\ket{x}) = \Phi(\ket{x_{\rm center}}) + \beta_{\rm J} \sum_{(i,j)\notin{\rm center}} S^z_i S^z_j\,,\label{eq:guide}
\end{align}
where $\Phi$ is a lookup table that maps $\ket{x_{\rm center}}$ to a real number, and $\beta_{\rm J}$ is a Jastrow coefficient, with the sum running over all edges that are not part of the central structure.
We use $\beta_{\rm J}=-3.5,\,-3.5,\,-3.0$ for $J_\pm/J_z=-0.025,\,-0.035,\,-0.045$, respectively.
The lookup table consists of $2^8=256$ parameters but was constructed to respect several spatial and spin symmetries of the central structure.
First, there are three independent automorphisms that interchange the spins on the apex bonds: $(1,2)$, $(4,5)$, and $(7,8)$ (see \autoref{fig:patch}).
Second, there is a mirror symmetry that exchanges the two outer triangles: $(1,8)(2,7)(3,6)$.
Third, in the absence of local fields, the Hamiltonian is invariant under a global spin flip.
These symmetries reduce the $256$ entries of the lookup table to only $33$ independent parameters.
We determine the parameters variationally by minimizing the energy using variational Monte Carlo within NetKet~\cite{netket2:2019,netket3:2022}.
The final parameters for each orbit, represented by a spin configuration, are listed in \autoref{tab:guide_orbits}.
Each of the $256$ spin configurations is represented by one of the $33$ states in the table.
For example, $\ket{x_{\rm center}}=\ket{10000000}$ has the same value as $\ket{01000000}$ because the two configurations are related by the automorphism that exchanges sites $1$ and $2$.
Similarly, $\ket{x_{\rm center}}=\ket{00000000}$ and $\ket{11111111}$ have the same value because they are related by a global spin flip.

\begin{table}[t]
\centering
\scriptsize
\setlength{\tabcolsep}{2pt}
\begin{tabular}{@{}ll@{\hspace{0.8em}}ll@{\hspace{0.8em}}ll@{}}
\hline\hline
state & $\Phi$ & state & $\Phi$ & state & $\Phi$ \\
\hline
\multicolumn{6}{c}{$J_{\pm}/J_z = -0.025$} \\
\hline
  $\ket{00000000}$ & $-0.9819$ & $\ket{11110000}$ & $-0.9063$ & $\ket{11010100}$ & $+2.4551$ \\
  $\ket{10000000}$ & $-1.2283$ & $\ket{00011000}$ & $-0.5128$ & $\ket{00110100}$ & $+0.2851$ \\
  $\ket{11000000}$ & $-1.0779$ & $\ket{10011000}$ & $-0.3734$ & $\ket{10110100}$ & $+0.0115$ \\
  $\ket{00100000}$ & $-0.7190$ & $\ket{11011000}$ & $-0.4132$ & $\ket{10011100}$ & $+0.3218$ \\
  $\ket{10100000}$ & $-1.0115$ & $\ket{00111000}$ & $-0.5064$ & $\ket{00111100}$ & $-0.5412$ \\
  $\ket{11100000}$ & $-0.8732$ & $\ket{10111000}$ & $-0.8744$ & $\ket{10111100}$ & $-1.0953$ \\
  $\ket{00010000}$ & $-0.9297$ & $\ket{10000100}$ & $+0.0116$ & $\ket{10000010}$ & $-1.1109$ \\
  $\ket{10010000}$ & $-1.1492$ & $\ket{11000100}$ & $+0.3477$ & $\ket{10100010}$ & $+0.0601$ \\
  $\ket{11010000}$ & $-0.9565$ & $\ket{00100100}$ & $+1.5909$ & $\ket{10010010}$ & $-0.1235$ \\
  $\ket{00110000}$ & $-0.3293$ & $\ket{10100100}$ & $+2.5190$ & $\ket{10110010}$ & $+2.9850$ \\
  $\ket{10110000}$ & $-0.4616$ & $\ket{10010100}$ & $+2.9003$ & $\ket{10011010}$ & $+2.6872$ \\
\hline
\multicolumn{6}{c}{$J_{\pm}/J_z = -0.035$} \\
\hline
  $\ket{00000000}$ & $-0.8738$ & $\ket{11110000}$ & $-0.9417$ & $\ket{11010100}$ & $+2.5608$ \\
  $\ket{10000000}$ & $-1.3621$ & $\ket{00011000}$ & $-0.6278$ & $\ket{00110100}$ & $+0.3861$ \\
  $\ket{11000000}$ & $-1.3337$ & $\ket{10011000}$ & $-0.2080$ & $\ket{10110100}$ & $+0.2663$ \\
  $\ket{00100000}$ & $-0.7858$ & $\ket{11011000}$ & $-0.2455$ & $\ket{10011100}$ & $+0.3790$ \\
  $\ket{10100000}$ & $-1.0314$ & $\ket{00111000}$ & $-0.6784$ & $\ket{00111100}$ & $-0.6326$ \\
  $\ket{11100000}$ & $-1.1714$ & $\ket{10111000}$ & $-0.8503$ & $\ket{10111100}$ & $-1.1146$ \\
  $\ket{00010000}$ & $-0.8588$ & $\ket{10000100}$ & $+0.2006$ & $\ket{10000010}$ & $-1.0732$ \\
  $\ket{10010000}$ & $-1.0779$ & $\ket{11000100}$ & $+0.3440$ & $\ket{10100010}$ & $+0.3157$ \\
  $\ket{11010000}$ & $-1.0484$ & $\ket{00100100}$ & $+1.4553$ & $\ket{10010010}$ & $+0.0946$ \\
  $\ket{00110000}$ & $-0.2476$ & $\ket{10100100}$ & $+2.3437$ & $\ket{10110010}$ & $+2.8389$ \\
  $\ket{10110000}$ & $-0.1129$ & $\ket{10010100}$ & $+2.5463$ & $\ket{10011010}$ & $+2.5446$ \\
\hline
\multicolumn{6}{c}{$J_{\pm}/J_z = -0.045$} \\
\hline
  $\ket{00000000}$ & $-1.1071$ & $\ket{11110000}$ & $-0.9050$ & $\ket{11010100}$ & $+2.2781$ \\
  $\ket{10000000}$ & $-1.4739$ & $\ket{00011000}$ & $-0.6135$ & $\ket{00110100}$ & $+0.5597$ \\
  $\ket{11000000}$ & $-1.2353$ & $\ket{10011000}$ & $-0.0943$ & $\ket{10110100}$ & $+0.5133$ \\
  $\ket{00100000}$ & $-0.8311$ & $\ket{11011000}$ & $-0.1369$ & $\ket{10011100}$ & $+0.5521$ \\
  $\ket{10100000}$ & $-1.0040$ & $\ket{00111000}$ & $-0.7354$ & $\ket{00111100}$ & $-0.7585$ \\
  $\ket{11100000}$ & $-1.2664$ & $\ket{10111000}$ & $-0.9615$ & $\ket{10111100}$ & $-1.1807$ \\
  $\ket{00010000}$ & $-1.1660$ & $\ket{10000100}$ & $+0.4795$ & $\ket{10000010}$ & $-1.3128$ \\
  $\ket{10010000}$ & $-1.2707$ & $\ket{11000100}$ & $+0.7146$ & $\ket{10100010}$ & $+0.5836$ \\
  $\ket{11010000}$ & $-1.1047$ & $\ket{00100100}$ & $+2.0996$ & $\ket{10010010}$ & $+0.2239$ \\
  $\ket{00110000}$ & $-0.1953$ & $\ket{10100100}$ & $+2.2594$ & $\ket{10110010}$ & $+2.4009$ \\
  $\ket{10110000}$ & $+0.2033$ & $\ket{10010100}$ & $+2.2231$ & $\ket{10011010}$ & $+2.2619$ \\
\hline\hline
\end{tabular}
\caption{The 33 independent coefficients of the lookup table used in \autoref{eq:guide} for the three different coupling strengths considered.
Each of the $2^8=256$ configurations of $\ket{x_{\rm center}}$ can be mapped to one of the representative states listed in the table using the automorphisms described in the text or the global spin flip.
}
\label{tab:guide_orbits}
\end{table}

\section{Detailed Geometry and Percolation Transitions}
In the main text, we discuss several important percolation transitions for different kinds of objects in the diluted quantum spin ice. Here, we
give precise definitions of the kinds of percolation discussed in the main text, and present the finite-size scaling analysis used to identify the percolation transitions. 

\begin{figure}
    \centering
    \includegraphics[width=\columnwidth]{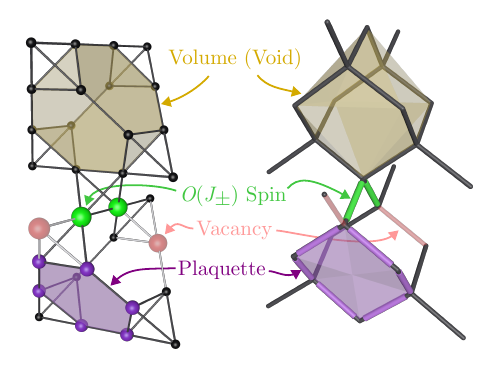}
    \caption{Dual-lattice representation of the diluted pyrochlore, showing the correspondence between pyrochlore sites (spheres, left) and links on the diamond lattice (graph edges, right). Spin vacancies are shown in pink. The two spins making up a length-2 string on which $O(J_\pm)$ processes act are shown in green.}
    \label{fig:diamond}
\end{figure}

First, we introduce some nomenclature. When discussing perturbation theory in spin ice, it is more convenient to view the lattice sites as lying on the medial (sometimes called ``dual'') lattice of the diamond lattice. The tetrahedron centers on which the emergent monopoles are supported lie on the diamond sites, as sketched in \autoref{fig:diamond}, while the spins are now understood as diamond links. The triangles hosting the bound monopoles are thus recast as vertices of order 3 on the diamond lattice. 

At higher levels of dilution, one may also end up with vertices of order 2 or even 1. In the former case, the ``ice rule'' simply locks the spins antiparallel to one another, while in the latter, we simply have a free spin that binds a monopole. These objects are, however, parametrically rarer than triangles in the low disorder limit. For clarity, we will therefore focus on only the tetrahedra and triangles in the discussion.

The perturbatively allowed processes within the ice manifold of clean spin ice must leave the charge state of each tetrahedron at zero. This means that the Dirac string that transports the virtual monopole must form a closed loop, producing the ring exchange. In our case, strings with open ends can now enter the effective Hamiltonian so long as these strings terminate at triangles. 

The $S_i^+S^-_j$ processes of strength $J_\pm$ discussed in the main text then correspond precisely to strings of length 2 on the diamond lattice; the $O(J_\pm^2/J_z)$ processes correspond to length-4 strings. At the order comparable to ring exchange $O(J_\pm^3/J_z^2)$, we must also consider open length-6 strings.

Our main geometric calculation can be summarized as follows:

\begin{enumerate}
\itemsep0em
    \item Generate a periodic pyrochlore lattice of $L^3$ 16-site cubic unit cells. 
    \item Flag spin sites as `deleted' with probability $p$.
    \item Identify all order-3 vertices on the dual lattice, i.e. triangles. These are the only points at which an open Dirac string may begin or end.
    \item For each spin (link), compute a list of diamond-graph distances to all neighboring triangles. A given triangle is called a spin's 0-neighbor if the spin lies on the triangle, 1-neighbor if the triangle lies one link away, and so on.
    \item Identify which strings a given spin is part of using these local lists. An $O(J_\pm)$ spin has a 0-neighbor and a 1-neighbor, while a $O(J_\pm^2/J_{z})$ spin has either a 0- and a 3-neighbor or a 1- and a 2-neighbor.
    \item Mark all spins not tied up in lower-order processes as non-defect spins.
    \item Mark all hexagons consisting of non-defect spins as ringflip-only hexagons.
    \item Mark all closed volumes of four ringflip-only hexagons as ringflip-only volumes.
\end{enumerate}

Once all spins have been appropriately classified, we identify connected clusters using a standard union-find algorithm. We consider a cluster to percolate if it winds around any of the three periodic boundaries.

\begin{figure}
    \centering
    \includegraphics[width=\linewidth]{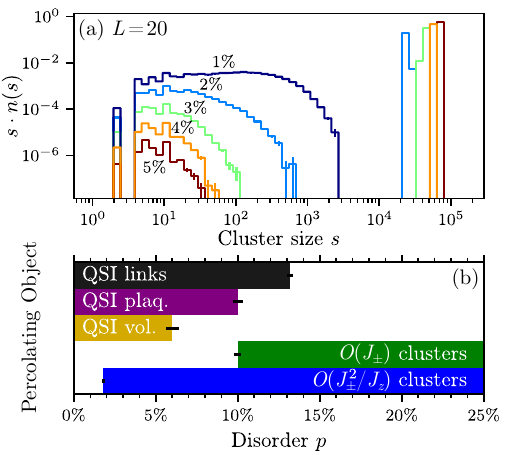}
    \caption{Fine-grained picture of percolation transitions in diluted QSI. Panel (a) shows the size distribution $s\cdot n(s)$ of clusters governed by $O(J_\pm^2/J_z)$ physics. We use a cubic system with $L=20$, consisting of approximately $16L^3(1-p)$ spins.  
    Panel (b) shows five distinct percolation transitions for different geometric objects. For the bottom two rows, we compute the clusters of spins connected by $O(J_\pm^2/J_{z})$ or $O(J_\pm)$ processes and check if they percolate; for the top three, we remove the $O(J_\pm^2/J_z)$ spins from the system and check whether the links, plaquettes or voids (see \autoref{fig:diamond}) percolate. All error bars and transition points are extracted from finite-size scaling on the range $L\in [20,100]$. 
    }
    \label{fig:cluster_dist}
\end{figure}

The cluster size distribution of $O(J_\pm^2/J_z)$ spins (i.e. spins which are part of a 4-string or 2-string process) on a $L=20$ cluster is given in \figref{fig:cluster_dist}{a}. It is plain to see that the many small clusters quickly assemble into one large cluster with size set by the system volume, even at $p\simeq 0.02$.

We evaluated percolation probabilities for a range of system sizes $L$ between 20 and 100. This allowed us to reliably extract the $L\to\infty$ percolation threshold by performing a finite-size scaling collapse. 

The lattice gauge theory of clean QSI naturally gives rise to three important classes of geometric objects---the diamond links, plaquettes and volumes, corresponding exactly to spins, plaquettes and `voids' (sometimes called `dual tetrahedra'). Respectively, these are the smallest objects on which one can define the electric field, photon fluctuations, and visons (Dirac monopoles). It can be plainly seen in~\figref{fig:cluster_dist}{b} that volume percolation is significantly more fragile to disorder than both plaquette and link percolation. 

\section{Hybrid Quantum-Classical Monte Carlo (ED+MC)}

\begin{figure}
    \centering
    \includegraphics[width=0.7\linewidth]{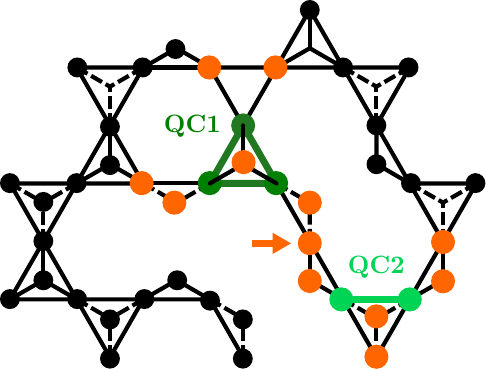}
    \caption{Illustration of two quantum clusters, QC1 and QC2, as used for ED+MC simulations. The $O(J_\pm)$ processes allow only the quantum spins (green circles) to fluctuate; each active $S^+_iS^-_j$ term is represented as a thick, green bond. All other bonds contain the longitudinal term only. The orange circles are the classical `boundary spins', coupled to the quantum clusters via Ising $S^zS^z$ interactions. The boundary spin touching both clusters (indicated with an arrow) cannot participate in any off-diagonal processes: all of its 1-neighbor diamond vertices are intact. Black circles indicate purely classical spins. Defect sites are not shown.}
    \label{fig:qc_def}
\end{figure}

We now discuss the hybrid technique we used in generating the heat capacity and neutron-scattering structure factors of the first-order Hamiltonian which does not percolate at the considered 5\% dilution.
Therefore, quantum effects are confined to finite, disconnected clusters.
The first-order effective Hamiltonian $H_{\rm eff}^{(1)}$ is a simple correction term to classical spin ice,
\begin{align}
    H = \sum_{\langle ij \rangle} J_z S^{z}_{i} S^{z}_{j} + \sum_{\langle ij \rangle \in \text{2-strings} } J_\pm (S^{+}_{i}S^{-}_{j}+S^{-}_{i}S^{+}_{j}).
    \label{S_eq:heff1}
\end{align}
We call the spins participating in 2-strings (i.e. the green spins in \autoref{fig:qc_def}) `quantum spins,' while the remainder are called `classical'. 
The Monte Carlo state vector then consists of two parts: a bit string containing the state of all classical spins, and a vector of eigenvalue indices specifying the eigenstate of each quantum cluster. 

The classical spins bordering the quantum clusters (see \autoref{fig:qc_def}) are treated as static boundary conditions for each cluster diagonalization.
Whenever a Metropolis proposal attempts to flip one of these spins, we re-diagonalize the system and resample its eigenvalue index on a Boltzmann distribution. We find that these moves, in addition to local spin flips and ring flips, are sufficient to give good convergence as monitored by acceptance rate and autocorrelation. All results presented in the main text are obtained by Metropolis--Hastings simulated annealing, beginning at $T=100J_z$ and cooling to $T=0.01J_{z}$ via 100 logarithmically spaced temperature steps. We perform 1024 sweeps at each temperature, and average over 64 disorder realizations.

Under these definitions, it is possible (though rare) for two distinct quantum clusters to share a `classical' $S^zS^z$ bond without any off-diagonal matrix element. In this case, we approximate the cluster-cluster interactions at mean-field level.

\begin{figure}
    \centering
    \includegraphics[width=\linewidth]{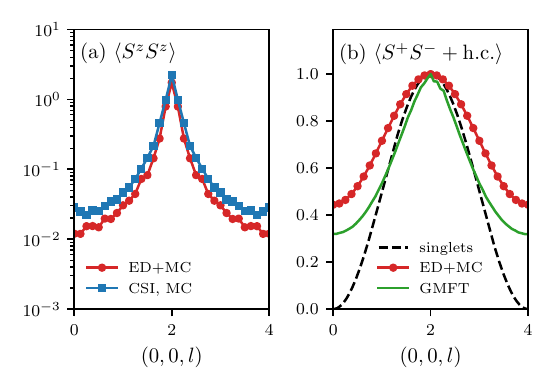}
    \caption{Cuts of various correlation functions along the $(0,0,l)$ line in momentum space. 
    Panel (a) compares $\langle S^z(q)S^z(-q)\rangle$ obtained via the hybrid ED+MC protocol with pure classical Monte Carlo on a diluted lattice, in each case at $T=0.01J_{z}$. For ED+MC, we take $J_\pm=0.2J_{z}$.
    Panel (b) shows a cut of \figref{fig:3}{b}, comparing finite-temperature ED+MC to zero-temperature singlets and GMFT, again at $J_\pm=0.2J_{z}$. Each curve is normalized such that they overlap at (0,0,2).}
    \label{fig:supp_cuts}
\end{figure}

The line cuts shown in \autoref{fig:supp_cuts} complement \autoref{fig:3} of the main text. In the diagonal channel $\langle S^zS^z \rangle$, we see that correlations are largely unaffected by the formation of dimers since dimers are being formed within the ice manifold. In \figref{fig:supp_cuts}{b}, however, we see that the clean-lattice, long-range entangled GMFT result can only be distinguished from simple local dimers by a subtle difference in curvature near the pinch point. This once again underscores the need to carefully characterize disorder in future experimental searches for spin liquid physics.

\end{document}